\documentclass[]{aa}
\usepackage{graphicx}
\usepackage{color}
\usepackage{txfonts}

\begin{document}

\title{Damping of Alfv\'en waves in solar partially ionized plasmas: effect of neutral helium in multi-fluid approach}

\author{Zaqarashvili, T.V.\inst{1,2}, Khodachenko, M.L. \inst{1} and Rucker, H.O. \inst{1}
}

 \institute{Space Research Institute, Austrian Academy of Sciences, Schmiedlstrasse 6, 8042 Graz, Austria\\
             \email{[teimuraz.zaqarashvili;maxim.khodachenko;rucker]@oeaw.ac.at}
                               \and
            Abastumani Astrophysical Observatory at Ilia State University, University St. 2, Tbilisi, Georgia\\
}

\date{Received / Accepted }

\abstract{Chromospheric and prominence plasmas contain neutral atoms, which may change the plasma dynamics through collision with ions. Most of the atoms are neutral hydrogen, but a significant amount of neutral helium may also be present in the plasma with a particular temperature. Damping of MHD waves due to ion collision with neutral hydrogen is well studied, but the effects of neutral helium are largely unknown.}{We aim to study the effect of neutral helium in the damping of Alfv\'en waves in solar partially ionized plasmas.}{We consider three-fluid magnetohydrodynamic (MHD) approximation, where one component is electron-proton-singly ionized helium and other two components are the neutral hydrogen and neutral helium atoms. We derive the dispersion relation of linear Alfv\'en waves in isothermal and homogeneous plasma. Then we solve the dispersion relation and derive the damping rates of Alfv\'en waves for different plasma parameters.}{The presence of neutral helium significantly enhances the damping of Alfv\'en waves compared to the damping due to neutral hydrogen at certain values of plasma temperature ($10000-40000$ K) and ionization. Damping rates have a peak near the ion-neutral collision frequency, but decrease for the higher part of wave spectrum.}{Collision of ions with neutral helium atoms can be of importance for the damping of Alfv\'en waves in chromospheric spicules and in prominence-corona transition regions.}


\titlerunning{Damping of Alfv\'en waves in partially ionized plasma}

\authorrunning{Zaqarashvili et al.}

\maketitle

\section{Introduction}

Existence of neutral atoms may change the dynamics of astrophysical plasma due to their collision with charged particles. Ion-neutral collision may lead to the damping of magnetohydrodynamic (MHD) waves in the solar photosphere/chromosphere and prominences  (Khodachenko el al. \cite{Khodachenko2004}, Forteza et al. \cite{Forteza2007}, Zaqarashvili et al. \cite{Zaqarashvili2011}). Only part of hydrogen atoms are ionized in the solar photosphere, but the ionization degree increases with height due to the rise of plasma temperature. The plasma in the transition region/lower corona is almost fully ionized, but upper chromosphere, spicules and prominences contain significant amount of neutral atoms. Numerous papers study the effects of plasma partial ionization in different aspects of the solar atmosphere (Haerendel \cite{Haerendel1992}, De Pontieu and Haerendel \cite{De Pontieu1998}, De Pontieu et al. \cite{De Pontieu2001}, Khodachenko and Zaitsev \cite{Khodachenko2002}, James and Erd\'elyi \cite{James2002}, James et al. \cite{James2004}, Khodachenko el al. \cite{Khodachenko2004}, Khodachenko and Rucker \cite{Khodachenko2005}, Leake et al. \cite{Leake2005}, Leake and Arber \cite{Leake2006}, Arber et al. {\cite{Arber2007}}, Forteza et al. \cite{Forteza2007,Forteza2008}, Vranjes et al. \cite{Vranjes2008}, Soler et al. \cite{Soler2009a,Soler2009b}, Gogoberidze et al. \cite{Gogoberidze2009}, Carbonell et al. \cite{Carbonell2010}, Singh and Krishan \cite{Singh}, Goodman \& Kazeminezhad \cite{Goodman2010}, Tsap et al. \cite{Tsap2011}). Most of these papers consider only the effects of neutral hydrogen and use a single-fluid MHD approximation. Recently, Zaqarashvili et al. (\cite{Zaqarashvili2011}) showed that the consideration of two-fluid approach significantly change the dynamics of MHD waves compared to the single-fluid one when wave frequency is comparable or higher than ion-neutral collision frequency. They show that the damping rates of transverse waves (Alfv\'en, fast magneto-acoustic) due to ion-neutral collision reach maximum value at certain wavelength, but then begin to decrease for higher frequencies. The two fluid-description shed light on few disputed problems, such as, for example, the damping of slow magneto-acoustic waves (Forteza et al. \cite{Forteza2007}). Therefore, the multi-fluid approach reveals more complete dynamics of partially ionized plasmas.

On the other hand, hydrogen is not the only sort of neutral atoms, which may be important in the damping of MHD waves. The solar plasma may contain significant amount of neutral helium in certain regions of the solar atmosphere. The upper chromosphere/transition region, spicules and prominences are the regions, where the presence of neutral helium may be of potential importance for the damping of MHD waves. The first step towards the inclusion of neutral helium has been done by Soler et al. (\cite{Soler2010}). They used the single-fluid MHD approach and concluded that the neutral helium causes no significant influence on the damping of MHD waves in the prominence plasma. However, the effect of neutral helium can be enhanced in the regions of the solar atmosphere with higher temperature, therefore the further analysis is of importance.

In this paper, we study the Alfv\'en waves in multi-fluid partially ionized plasma, which contain neutral hydrogen and neutral helium atoms. The particular attention is paid to the wave damping due to ion-neutral
collisions and the difference between damping rates in partially ionized plasma with and without neutral helium. We derive the three-fluid MHD equations from initial five-fluid equations and solve the linearized equations in the simplest case of a homogeneous plasma.

\section{Main equations}

We study partially ionized plasma, which consists in
electrons (${\mathrm {e}}$), protons (${\mathrm {H^+}}$), singly ionized helium (${\mathrm {He^+}}$), neutral hydrogen (${\mathrm {H}}$) and neutral helium (${\mathrm {He}}$). We suppose that each sort of
species has Maxwell velocity distribution, therefore they can be described as separate fluids. Below we first write the equations in five-fluid description and then perform consequent transition to three-fluid approach.

\subsection{Multi-fluid equations}

We aim to study the dynamics of Alfv\'en waves, therefore we consider the incompressible plasma. We also neglect the viscosity, the heat flux and the heat production due to collisions between particles. Then the fluid equations for each species can be written as (Braginskii
\cite{Braginskii1965}, Goedbloed \& Poedts \cite{Goedbloed2004})

\begin{equation}\label{ne3}
\nabla \cdot \vec V_a=0,
\end{equation}
$$
m_an_a\left ({{\partial \vec V_a}\over {\partial t}}+({\vec
V_a}\cdot \nabla)\vec V_a\right )=-\nabla p_{a}-e_a n_a\left (\vec E +{1\over c}\vec V_a \times \vec B \right )+
$$
\begin{equation}\label{Ve3}
+\vec
R_a,
\end{equation}
where $m_a$, $n_a$, $p_a$, $\vec V_a$ are the mass, the number
density, the pressure, and the velocity of particles
$a$, $\vec E$ is the electric field, $\vec B $ is the magnetic field
strength, $\vec R_a$ is the change of impulse of particles $a$ due to
collisions with other sort of particles, $e_a=\pm 4.8\times 10^{-10}$ statcoul is the charge of electrons, protons and singly ionized helium (note, that $e_a=0$ for neutral particles) and
$c=2.9979\times 10^{10}$ cm s$^{-1}$ is the speed of light. Plasma is supposed to be quasi-neutral, which means
$n_{\mathrm {e}}=n_{\mathrm {H^+}}+n_{{\mathrm {He^+}}}$. The description of the system is completed by Maxwell
equations which have the forms (without displacement current)
\begin{equation}\label{e}
\nabla \times \vec E=-{1\over c}{{\partial \vec B}\over {\partial t}},
\end{equation}
\begin{equation}\label{B}
\nabla \times \vec B={{4 \pi} \over c}{\vec j},
\end{equation}
where
\begin{equation}\label{j2}
\vec j=-e(n_{\mathrm {e}}\vec V_{\mathrm {e}}-n_{\mathrm {H^+}}\vec V_{{\mathrm {H^+}}}-n_{{\mathrm {He^+}}}\vec V_{{\mathrm {He^+}}})
\end{equation}
is the current density.

In the case of Maxwell distribution in each sort of particles, $\vec
R_a$ has the form (Braginskii \cite{Braginskii1965}):
\begin{equation}\label{Re}
\vec R_a=-\sum_b \alpha_{ab}(\vec V_a-\vec V_b),
\end{equation}
where $\alpha_{ab}=\alpha_{ba}$ are coefficients of friction between particles $a$ and $b$.

For time scales longer than ion-electron and ion-ion collision times, the
electron and ion gases can be considered as a single fluid. This
significantly simplifies the equations taking into account the
smallness of electron mass with regards to the masses of ion and
neutral atoms. Then the five-fluid description can be changed by
three-fluid description, where one component is the charged fluid (electron+protons+singly ionized helium) and the other two components are the gases of neutral hydrogen and neutral helium.

We use the definition of
total density and the total velocity of charged fluid as
\begin{equation}\label{rho}
\rho=\rho_{{\mathrm {H^+}}}+\rho_{{{\mathrm {He^+}}}},
\end{equation}
\begin{equation}\label{V1}
{{\vec V}}= {{\rho_{{\mathrm {H^+}}}\vec V_{\mathrm {H^+}}+\rho_{{\mathrm {He^+}}}\vec V_{{\mathrm {He^+}}}}\over
{\rho_{{\mathrm {H^+}}}+\rho_{{\mathrm {He^+}}}}},
\end{equation}
correspondingly. Then the sum of momentum equations for electron, proton and singly ionized helium gives the equation
\begin{equation}\label{V}
\rho{{d {{\vec V}}}\over {d t}}+ \rho \xi_{{\mathrm {H^+}}} \xi_{\mathrm {He^+}} (\vec w \nabla) \vec w =-\nabla p_{}+{1\over c}\vec j \times \vec B + \vec F_t,
\end{equation}
where
$$
\vec F_t=-(\alpha_{{\mathrm {H^+H}}}+\alpha_{{\mathrm {H^+He}}}+\alpha_{{\mathrm {He^+H}}}+\alpha_{{\mathrm {He^+He}}}){{\vec V}}+
$$
$$
+\xi_{{\mathrm {H^+}}}(\alpha_{{\mathrm {He^+He}}}+\alpha_{{\mathrm {He^+H}}})\vec w-\xi_{{\mathrm {He^+}}}(\alpha_{{\mathrm {H^+H}}}+\alpha_{{\mathrm {H^+He}}})\vec w+
$$
\begin{equation}\label{F}
+(\alpha_{{\mathrm {H^+H}}}+\alpha_{{\mathrm {He^+H}}})\vec V_{{\mathrm {H}}}+(\alpha_{{\mathrm {H^+He}}}+\alpha_{{\mathrm {He^+He}}})\vec V_{{\mathrm {He}}}
\end{equation}
is the collision term, $p=p_{\mathrm {e}}+p_{\mathrm {H^+}}+p_{\mathrm {He^+}}$ is the total pressure of charged fluid,
$\xi_{{\mathrm {H^+}}}={{m_{{\mathrm {H^+}}}n_{{\mathrm {H^+}}}}/\rho}$, $\xi_{\mathrm {He^+}}={{m_{{\mathrm {He^+}}}n_{{\mathrm {He^+}}}}/\rho}$
are relative concentrations of ions and $\vec w= \vec V_{\mathrm {H^+}} - \vec V_{{\mathrm {He^+}}}$ is the relative velocity of protons
and helium ions. In derivation of Eq. \ref{V} we neglect the electron inertia.
Friction of plasma with neutral atoms is mostly defined by the ion-neutral collision ($\alpha_{en}/\alpha_{in}\sim \sqrt{m_e/m_i}\ll 1$), therefore the collision of electrons with neutrals is also neglected here.

The relative velocity of heavy ions, $\vec w$ is small compared to the center of mass velocity. This can be easily seen if one uses the momentum equations for proton and singly ionized helium without neutral atoms. Subtracting the equations one may get
$$
(m_{\mathrm {H^+}}-m_{\mathrm {He^+}}){{d {{\vec V}}}\over {d t}}+ (m_{\mathrm {H^+}}\xi_{\mathrm {He^+}}+m_{\mathrm {He^+}}\xi_{{\mathrm {H^+}}}){{d {{\vec w}}}\over {d t}}+
$$
$$
+(m_{\mathrm {H^+}}\xi^2_{\mathrm {He^+}} - m_{\mathrm {He^+}}\xi^2_{{\mathrm {H^+}}})(\vec w \nabla) \vec w =
$$
\begin{equation}\label{w}
=-\left ({{\nabla p_{\mathrm {H^+}}}\over {n_{\mathrm {H^+}}}} -{{\nabla p_{\mathrm {He^+}}}\over {n_{\mathrm {He^+}}}}\right )+{e\over c}\vec w \times \vec B-\alpha_{{\mathrm {H^+He^+}}}\left ({1\over {n_{\mathrm {H^+}}}}+{1\over {n_{\mathrm {He^+}}}}
\right )\vec w.
\end{equation}
The ratio of collision and Lorentz terms is proportional to $\delta_{\mathrm {H^+He^+}}/\Omega_{\mathrm {H^+}}$, where $\Omega_{\mathrm {H^+}}=eB/m_{\mathrm {H^+}} c$ is the proton gyrofrequency ($\sim 10^5 s^{-1}$ for the magnetic field strength of 10 G) and $\delta_{\mathrm {H^+He^+}}$ is the ion-ion collision frequency. This ratio is much smaller than unity in the solar chromosphere, therefore the collision term can be neglected. The ratio of the second term in the left hand side and Lorentz term is proportional to $\omega/\Omega_{\mathrm {H^+}}$, where $\omega$ is the wave characteristic frequency. This ratio is much less than unity for the time scales much longer than ion gyroperiod, therefore the inertial term with $\vec w$ can be also neglected. The third term in the left hand side is even smaller. Then one may get
\begin{equation}\label{w1}
{w\over V}\sim {{\omega}\over {\Omega_{\mathrm {H^+}}}},
\end{equation}
which yields that $w\ll V$ for the time scales much longer than ion gyroperiod. Therefore, the terms in Eqs. \ref{V}-\ref{F}, which contain $\vec w$, are smaller comparing to other terms. Consideration of these terms complicate the calculations, but has no significant influence on the final results as we are interested in the effects of ion-neutral collisions. Therefore, the terms are neglected in the rest of the paper.

Momentum equations for neutral hydrogen and neutral helium atoms can be written as

\begin{equation}\label{VHI}
\rho_{{\mathrm {H}}}{{d \vec V_{{\mathrm {H}}}}\over {d t}} =-\nabla p_{{{\mathrm {H}}}}+\vec F_{{\mathrm {H}}},
\end{equation}
\begin{equation}\label{VHeI}
\rho_{{\mathrm {He}}}{{d \vec V_{{\mathrm {He}}}}\over {d t}} =-\nabla p_{{\mathrm {He}}}+\vec F_{{\mathrm {He}}},
\end{equation}
where the collision terms are
$$
\vec F_{{\mathrm {H}}}=-(\alpha_{{\mathrm {H^+H}}}+\alpha_{{\mathrm {He^+H}}}+\alpha_{{\mathrm {HeH}}})\vec V_{{\mathrm {H}}}+\alpha_{{\mathrm {HeH}}}\vec V_{{\mathrm {He}}}+
$$
\begin{equation}\label{FHI3}
+(\alpha_{{\mathrm {H^+H}}}+\alpha_{{\mathrm {He^+H}}}){{\vec V}},
\end{equation}
$$
\vec F_{{\mathrm {He}}}=-(\alpha_{{\mathrm {H^+He}}}+\alpha_{{\mathrm {He^+He}}}+\alpha_{{\mathrm {HeH}}})\vec V_{{\mathrm {He}}}+\alpha_{{\mathrm {HeH}}}\vec V_{{\mathrm {H}}}+
$$
\begin{equation}\label{VHeI3}
+(\alpha_{{\mathrm {H^+He}}}+\alpha_{{\mathrm {He^+He}}}){{\vec V}}.
\end{equation}

Ohm's law is obtained from the momentum equation of electrons and it takes the form
\begin{equation}\label{E}
\vec E= -{{\nabla p_{\mathrm {e}}}\over {en_{\mathrm {e}}}}+{1\over {cen_{\mathrm {e}}}}\vec j \times \vec B-{1\over {c}}{{\vec V}} \times \vec B+ {{\alpha_{{\mathrm {eH^+}}}+\alpha_{{\mathrm {eHe^+}}}}\over {e^2n^2_{\mathrm {e}}}}\vec j.
\end{equation}
Here the terms containing the ion relative velocity, $\vec w$, and electron-neutral collision are again neglected.

Maxwell equation (Eq. \ref{e}) and Ohm's law (Eq. \ref{E}) lead to
the induction equation
\begin{equation}\label{B2}
{{\partial \vec B}\over {\partial t}}={\nabla \times}({{\vec V}}
\times \vec B),
\end{equation}
where the battery, Hall and Ohmic diffusion terms are neglected.

\subsection{Ion-neutral collision frequency}

The coefficient of friction between ions and neutrals (in the case of same temperature) is calculated as (Braginskii \cite{Braginskii1965})
\begin{equation}\label{in}
{\alpha_{in}}= n_i n_n m_{in}\sigma_{in}{4\over 3} \sqrt{{{8kT}\over {\pi m_{in}}}}  .
\end{equation}
where $T$ is the plasma temperature, $m_i$ ($m_n$) is the ion (neutral atom) mass, $m_{in}=m_i m_n /(m_i+m_n)$ is reduced mass, $n_i$ ($n_n$) is the ion (neutral atom) number density, $\sigma_{in}=\pi (r_i+r_n)^2=4\pi
r^2_i$ is ion-neutral collision cross section  and
$k=1.38\times 10^{-16}$ erg K$^{-1}$ is the Boltzmann constant. Using mean atomic cross section  $\pi r^2_i=8.7974\times 10^{-17}$ cm$^{2}$ and the atomic masses of hydrogen and helium, the expression can be rewritten for each collision pairs as
$$
{\alpha_{{\mathrm {H^+H}}}}\approx 8\times 10^{-36} n_{{\mathrm {H^+}}} n_{{\mathrm {H}}} \sqrt{T} \,\, g \, cm^{-3} s^{-1},
$$
$$
{\alpha_{{\mathrm {H^+He}}}}\approx 10^{-35} n_{{\mathrm {H^+}}} n_{{\mathrm {He}}} \sqrt{T} \,\, g \, cm^{-3} s^{-1},
$$
$$
{\alpha_{{\mathrm {He^+H}}}}\approx 10^{-35} n_{{\mathrm {He^+}}} n_{{\mathrm {H}}} \sqrt{T} \,\, g \, cm^{-3} s^{-1},
$$
\begin{equation}\label{HHI}
{\alpha_{{\mathrm {He^+He}}}}\approx 1.6\times 10^{-35} n_{{\mathrm {He^+}}} n_{{\mathrm {He}}} \sqrt{T} \,\, g \, cm^{-3} s^{-1},
\end{equation}
where $T$ and $n$ are normalised by 1 $K$ and cm$^{-3}$, respectively.

\begin{figure}
\begin{center}
\includegraphics[width=8cm]{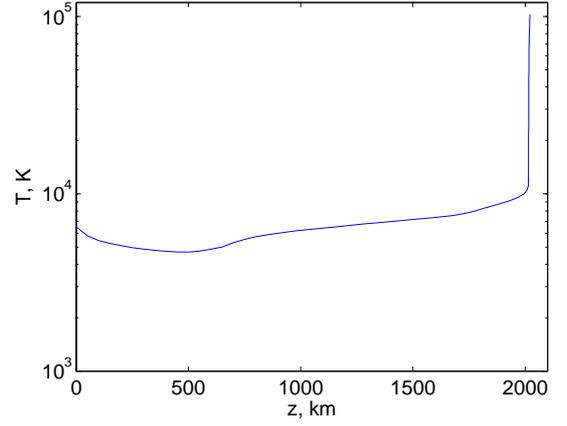}
\includegraphics[width=8cm]{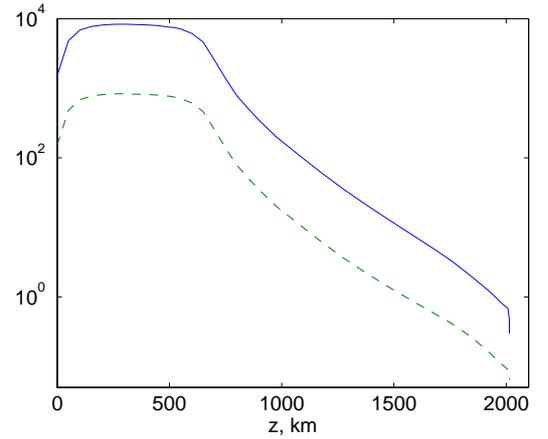}
\includegraphics[width=8cm]{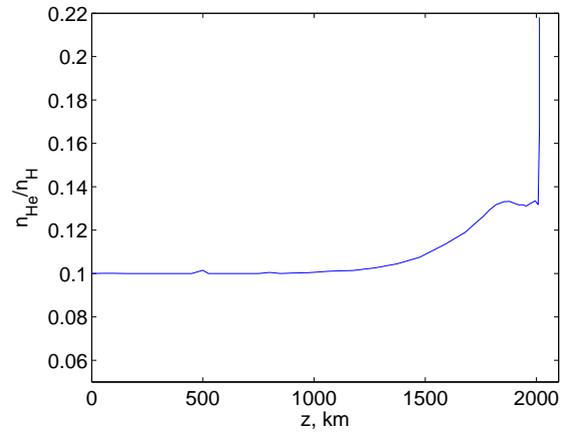}
\end{center}
\caption{Height dependence of atmospheric parameters according to FAL93-3 model (Fontenla et al. \cite{Fontenla1993}). Upper panel: the plasma temperature. Middle panel: blue solid line - the ratio of neutral hydrogen and electron number densities, $n_{\mathrm {H}}/n_{\mathrm {e}}$; green dashed line- the ratio of neutral helium and electron number densities, $n_{{\mathrm {He}}}/n_{\mathrm {e}}$. Lower panel: the ratio of neutral helium and neutral hydrogen number densities, $n_{{\mathrm {He}}}/n_{\mathrm {H}}$.}
\end{figure}


Collision frequency between ions and neutrals needs additional discussion. It is commonly accepted that the collision frequency of ions with neutrals $\nu_{in}=\alpha_{in}/{m_in_i}$ is generally different than the collision frequency of neutrals with ions $\nu_{ni}=\alpha_{in}/{m_nn_n}$
(De Pontieu et al. \cite{De Pontieu2001}, Vranjes et al. \cite{Vranjes2008}).
In the first (second) case the neutrals (ions) are supposed to have zero initial velocity and this approximation leads to the different values of 
collision frequency.
%
However, mean collision frequency between moving ions and neutral atoms is supposed to be the same value due to physical grounds. From simple equations of motions of ions and neutrals (without pressure and Lorentz terms) one can derive the equation for relative velocity between ions and neutrals
\begin{equation}\label{V_i-V_n}
{{\partial {{(\vec V_i-\vec V_n)}}}\over {\partial t}} =-\alpha_{in}\left ({1\over {m_i n_i}} + {1\over {m_n n_n}}\right )(\vec V_i - \vec V_n),
\end{equation}
where $\vec V_i$ and $\vec V_n$ are velocities of ions and neutral atoms, respectively. This equation gives the collision frequency between ions and neutrals as
\begin{equation}\label{nu_in}
\nu_{in}=\alpha_{in}\left ({1\over {m_i n_i}} + {1\over {m_n n_n}}\right ).
\end{equation}
This expression significantly changes the value of collision frequency between ions and neutrals. For example, calculation of collision frequency in the parameters used in Zaqarashvili et al. (\cite{Zaqarashvili2011}, Eq. 30) gives 16 s$^{-1}$ instead of 4 s$^{-1}$.

The chromospheric temperature of $1.6 \times 10^4$ K yields the ion-neutral collision frequencies (as expressed by Eq. \ref{nu_in}) as:
$$
\nu_{{\mathrm {H^+H}}}=\alpha_{{\mathrm {H^+H}}}\left ({{1}\over
{m_{\mathrm {H^+}}n_{{\mathrm {H^+}}}}}+ {{1}\over
{m_{{\mathrm {H}}}n_{{\mathrm {H}}}}}\right )\approx 61.4\,\, s^{-1},
$$
$$
\nu_{{\mathrm {H^+He}}}=\alpha_{{\mathrm {H^+He}}}\left ({{1}\over
{m_{\mathrm {H^+}}n_{{\mathrm {H^+}}}}}+ {{1}\over
{m_{{\mathrm {He}}}n_{{\mathrm {He}}}}}\right )\approx 19\,\, s^{-1},
$$
$$
\nu_{{\mathrm {He^+H}}}=\alpha_{{\mathrm {He^+H}}}\left ({{1}\over
{m_{\mathrm {H}}n_{{\mathrm {H}}}}}+ {{1}\over
{m_{{\mathrm {He^+}}}n_{{\mathrm {He^+}}}}}\right )\approx 7.2\,\, s^{-1},
$$
\begin{equation}\label{collision}
\nu_{{\mathrm {He^+He}}}=\alpha_{{\mathrm {He^+He}}}\left ({{1}\over
{m_{{\mathrm {He}}}n_{{\mathrm {He}}}}}+ {{1}\over
{m_{{\mathrm {He^+}}}n_{{\mathrm {H^+}}}}}\right ) \approx 2.8\,\, s^{-1},
\end{equation}
where we used proton, neutral hydrogen, singly ionized helium and neutral helium number densities of $9.09 \times 10^{10}$ cm$^{-3}$, $1.05 \times 10^{10}$ cm$^{-3}$, $6.9 \times 10^{9}$ cm$^{-3}$ and $2.5 \times 10^{9}$ cm$^{-3}$ (Fontenla et al. \cite{Fontenla1993}, model FAL93-3).

One may expect that the effects of ion-neutral collision may be enhanced for the waves with frequencies near ion-neutral collision frequency (Zaqarashvili et al. \cite{Zaqarashvili2011}). Therefore, higher frequency waves are probably more affected by ion-neutral collisions.


\subsection{Partial ionization in the solar chromosphere}

Before the study of damping of linear Alfv\'en waves due to ion-neutral collision, let us briefly discuss the plasma ionization in the solar lower atmosphere. The plasma is weakly ionized in the photosphere/lower chromosphere, but becomes more and more ionized with height. The increase of temperature with height leads to the ionization of hydrogen and helium atoms. Hydrogen and helium are almost fully ionized in the solar corona, so the transition occurs near the region of sharp temperature rise.

Figure 1 shows the dependence of plasma parameters on height according to FAL93-3 model (Fontenla et al. \cite{Fontenla1993}). This model includes the dependence of ionization degree on heights for both, hydrogen and helium. The upper panel shows the plasma temperature vs height according to this model. The temperature minimum is located near 500 km above the basic of the photosphere, while the transition region is just above 2000 km height. The neutral hydrogen number density is much higher than the electron number density at the lower heights, but becomes comparable near $\sim$ 1900 km, which corresponds to the temperature of 9400 K (blue solid line on the middle panel). Hydrogen atoms quickly become ionized above this height. The neutral helium number density is also higher than the electron number density at the lower heights (green dashed line on the middle panel). They become comparable near $\sim$ 1600 km, which corresponds to the temperature of $\sim$ 7300 K. The ratio becomes smaller and smaller just above this height. The important parameter for this study is the ratio of neutral helium and neutral hydrogen number densities (the lower panel). This ratio stays nearly constant ($\sim$ 0.1) up to 1500 km height, then it quickly increases up to 0.2 at height of 2000 km.
Therefore, the effect of neutral helium in the damping of Alfv\'en waves should become important at higher altitudes.

In the next section we study the damping of linear Alfv\'en waves due to ion collision with neutral hydrogen and helium atoms.

\section{Damping of linear Alfv\'en waves}

We consider the simplest case of a static and homogeneous plasma with
homogeneous unperturbed magnetic field, $B_z$, directed
along the $z$ axis. We assume the wave propagation in $xz$ plane i.e. $\partial
/\partial y=0$. Then the Alfv\'en waves are governed by the equations

\begin{equation}\label{Alfven-vl}
{{\partial v_y}\over {\partial t}}=
{B_z\over {4 \pi
\rho_{0}}}{{\partial b_{y}}\over {\partial z}} -{{\alpha_{{\mathrm {H}}}+\alpha_{{\mathrm {He}}}}\over \rho_{0}}v_y+{{\alpha_{{\mathrm {H}}}}\over \rho_{0}}v_{{\mathrm {H}}y}+{{\alpha_{{\mathrm {He}}}}\over \rho_{0}}v_{{\mathrm {He}}y},
\end{equation}
\begin{equation}\label{Alfven-uHl}
{{\partial v_{{\mathrm {H}}y}}\over {\partial t}}={{\alpha_{{\mathrm {H}}}}\over \rho_{H0}}v_y-{{\alpha_{{\mathrm {H}}}+\alpha_{{\mathrm {HeH}}}}\over \rho_{{\mathrm {H}}0}}v_{{\mathrm {H}}y}+{{\alpha_{{\mathrm {HeH}}}}\over \rho_{H0}}v_{{\mathrm {He}}y},
\end{equation}
\begin{equation}\label{Alfven-uHel}
{{\partial v_{{\mathrm {He}}y}}\over {\partial t}}={{\alpha_{{\mathrm {He}}}}\over \rho_{{\mathrm {He}}0}}v_y-{{\alpha_{{\mathrm {He}}}+\alpha_{{\mathrm {HeH}}}}\over \rho_{{\mathrm {He}}0}}v_{{\mathrm {He}}y}+{{\alpha_{{\mathrm {HeH}}}}\over \rho_{He0}}v_{{\mathrm {H}}y},
\end{equation}
\begin{equation}\label{Alfven-by}
{{\partial b_{y}}\over {\partial t}}=B_z{{\partial v_{y}}\over {\partial z}}.
\end{equation}
where  $v_y$ ($v_{{\mathrm {H}}y}$, $v_{{\mathrm {He}}y}$) are the perturbations of ion (neutral hydrogen, neutral helium) velocity, $b_y$ is the perturbation of the magnetic field, and $\rho_{0}$ ($\rho_{{\mathrm {H}}0}$, $\rho_{{\mathrm {He}}0}$) are unperturbed ion (neutral hydrogen, neutral helium) density, respectively. Here we use the definitions
\begin{equation}\label{a}
\alpha_{{\mathrm {H}}}=\alpha_{{\mathrm {H^+H}}}+\alpha_{{\mathrm {He^+H}}},\,\, \alpha_{{\mathrm {He}}}=\alpha_{{\mathrm {H^+He}}}+\alpha_{{\mathrm {He^+He}}}.
\end{equation}

%
\begin{figure}
\begin{center}
\includegraphics[width=8cm]{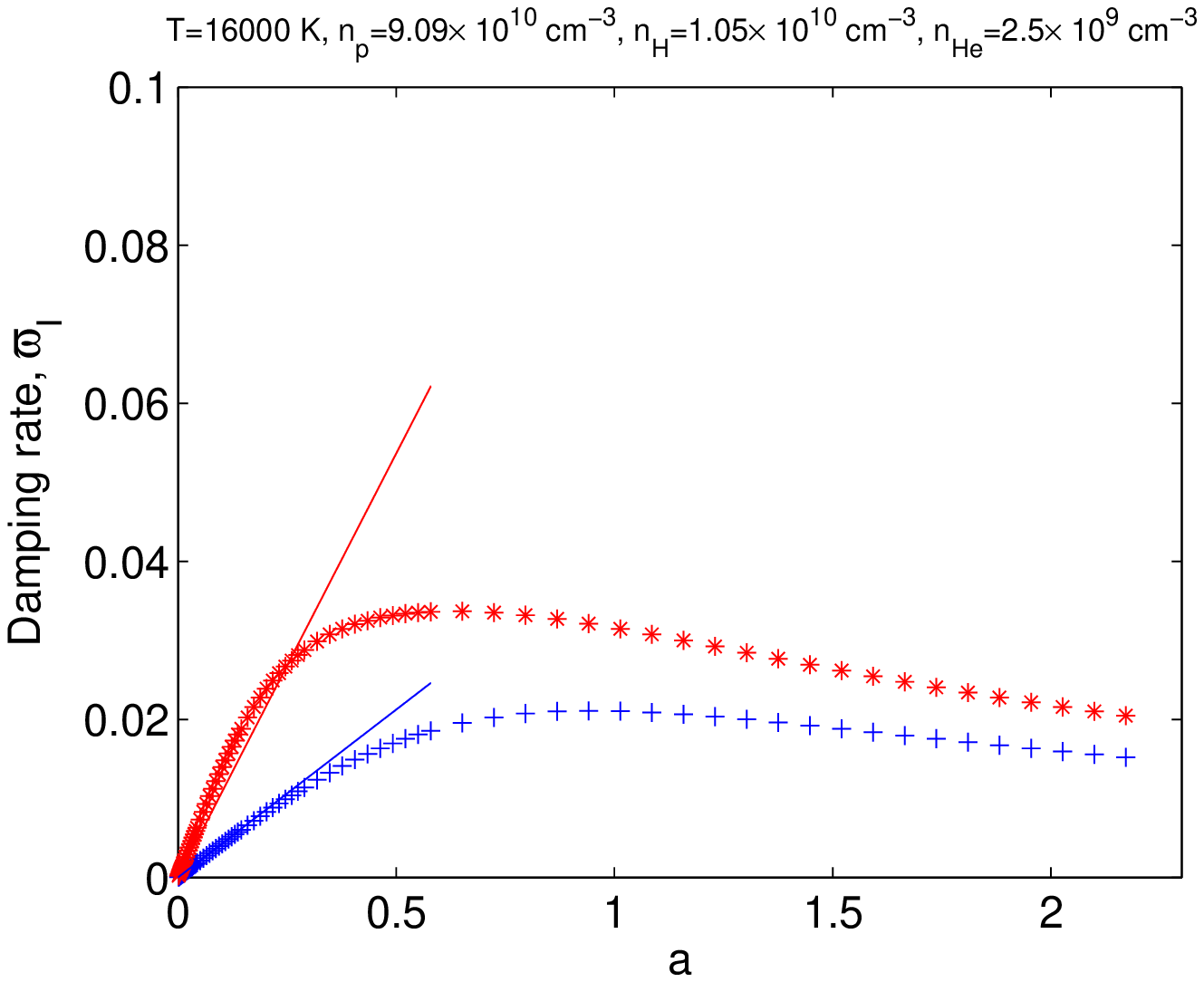}
\includegraphics[width=8cm]{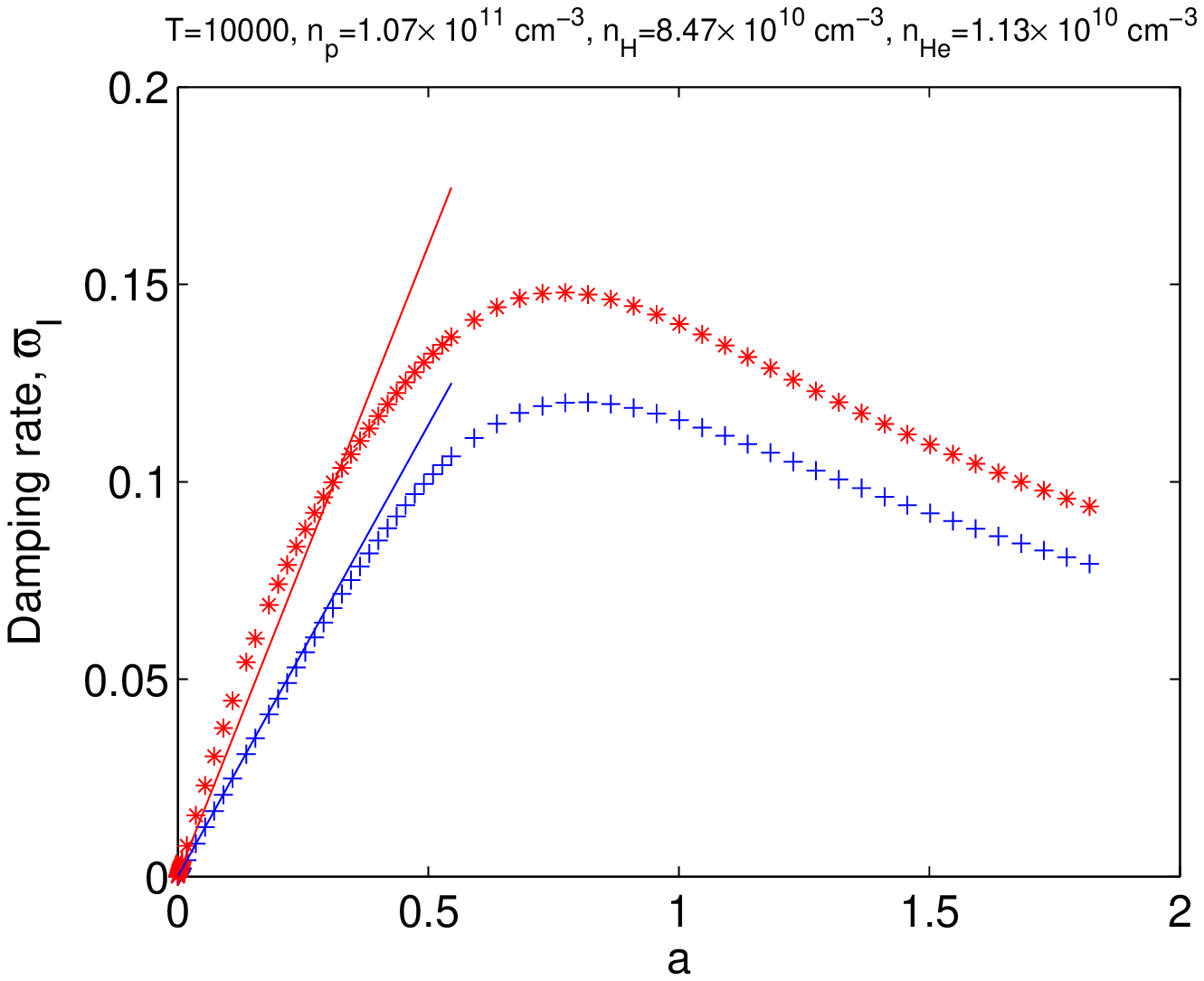}
\includegraphics[width=8cm]{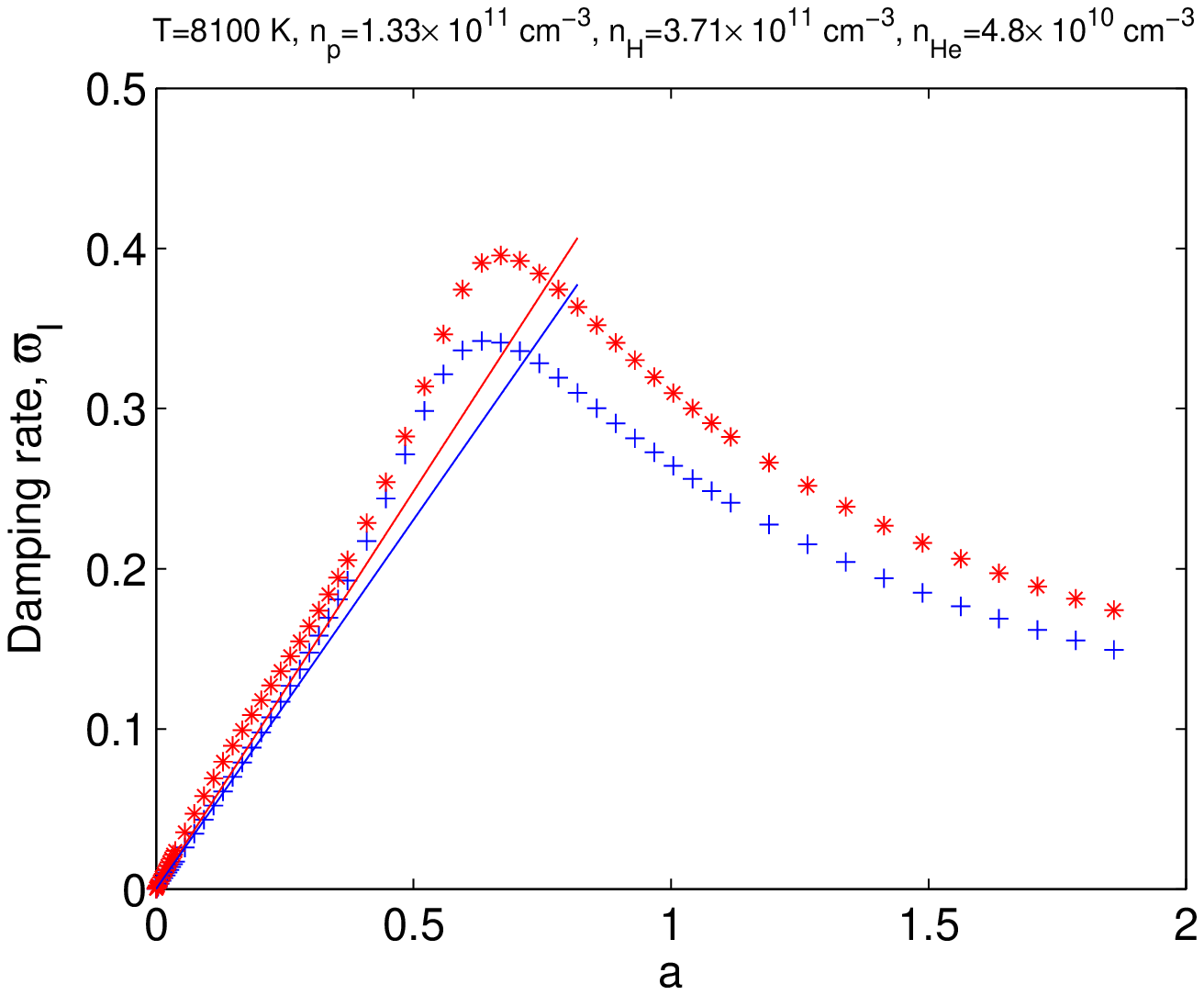}
\end{center}
\caption{Damping rate (imaginary part of frequency) of Alfv\'en waves, ${\varpi}={{\omega}/ {k_zv_A}}$, vs normalized Alfv\'en frequency, $a={{k_z
v_A}/ {\nu_{{\mathrm {H}}}}}$, for three deferent values of chromospheric temperature. Blue crosses correspond to the damping rates due to ion collision with neutral hydrogen atoms only, while red asterisks correspond to the damping rates due to ion collision with both, neutral hydrogen and neutral helium atoms. Red (blue) solid line corresponds to the damping rate derived in the single-fluid approach (Eq. \ref{single}) with (without) neutral helium.}
\end{figure}

Collision between neutral hydrogen and neutral helium atoms, expressed by $\alpha_{{\mathrm {HeH}}}$, does not contribute significantly in the damping of Alfv\'en waves. Moreover, for example, $\alpha_{{\mathrm {HeH}}}/\alpha_{{\mathrm {HeH^+}}}\sim n_{{\mathrm {H}}}/n_{{\mathrm {H^+}}}<1$ in spicules and in prominence plasma, therefore we neglect the corresponding terms for simplicity. Then, Fourier analysis with $exp[i(k_z z -\omega t)]$ gives the dispersion relation
$$
\xi_{{\mathrm {H}}}\xi_{{\mathrm {He}}}a_{{\mathrm {H}}}a_{{\mathrm {He}}}{\varpi}^4 +i[\xi_{{\mathrm {H}}}a_{{\mathrm {H}}}(1+\xi_{{\mathrm {He}}})+\xi_{{\mathrm {He}}}a_{{\mathrm {He}}}(1+\xi_{{\mathrm {H}}})]{\varpi}^3 -
$$
$$
-[\xi_{{\mathrm {He}}}+\xi_{{\mathrm {H}}}+1 +\xi_{{\mathrm {H}}}\xi_{{\mathrm {He}}}a_{{\mathrm {H}}}a_{{\mathrm {He}}}]{\varpi}^2 -i[\xi_{{\mathrm {H}}}a_{{\mathrm {H}}}+\xi_{{\mathrm {He}}}a_{{\mathrm {He}}}]{\varpi}+1= 
$$
\begin{equation}\label{disp-he}
=0,
\end{equation}
where
$$
{\varpi}={{\omega}\over {k_zv_A}},\,\,a_{{\mathrm {H}}}={{k_z v_A \rho_0}\over
{\alpha_{{\mathrm {H}}}}},\,\,a_{{\mathrm {He}}}={{k_z v_A \rho_0}\over
{\alpha_{{\mathrm {He}}}}},\,\, \xi_{{\mathrm {H}}}={\rho_{{\mathrm {H}}}\over \rho_0}, \,\, \xi_{{\mathrm {He}}}={\rho_{{\mathrm {He}}}\over \rho_0},
$$
\begin{equation}\label{alfven-speed}
v_A= {B_z\over {\sqrt{4 \pi \rho_{0}}}}.
\end{equation}
The dispersion relation has four different roots: the two complex solutions, which correspond to Alfv\'en waves damped by ion-neutral collision and two purely imaginary solutions, which correspond to damped vortex solutions of neutral hydrogen and neutral helium fluids.

In order to study the effects of neutral helium, the damping rates of Alfv\'en waves obtained from Eq. (\ref{disp-he}) should be compared with the damping rates obtained in the absence of neutral helium. The dispersion relation corresponding to Alfv\'en waves in two-fluid plasma with only neutral hydrogen can be easily obtained from Eq. (\ref{disp-he}) (see also Zaqarashvili et al. 2011)
\begin{equation}\label{disp-h}
\xi_{{\mathrm {H}}}a_{{\mathrm {H}}}{\varpi}^3 +i(1+\xi_{{\mathrm {H}}}){\varpi}^2 - \xi_{{\mathrm {H}}}a_{{\mathrm {H}}}{\varpi}-i=0.
\end{equation}
This dispersion relation has three different roots: the two complex solutions, which correspond to damped Alfv\'en waves and a purely imaginary solution. The purely imaginary solution is associated with the neutral fluid. This is clearly seen if one neglects the ion-neutral collision and consequently the neutral fluid behaves independently. Then, the fluid has vortex, $\omega=0$, solution in the incompressible limit, which gains imaginary part when collision is included. Therefore, the purely imaginary solution of dispersion relation Eq. (\ref{disp-h}) corresponds to damped vortex solution of neutral hydrogen fluid (Zaqarashvili et al. 2011). Then, it is straightforward that the fourth imaginary solution in Eq. (\ref{disp-he}) arises owing to the vortex solution of neutral helium fluid as mentioned above.

Fig. 2 displays the damping rates of Alfv\'en waves in partially ionized isothermal plasma with (red asterisks, derived from Eq.
\ref{disp-he}) and without (blue crosses, derived from Eq. \ref{disp-h}) neutral helium vs normalised Alfv\'en frequency for three different temperatures in the chromosphere. The Alfv\'en frequency $k_z v_A$ is normalised on the collision frequency between neutral hydrogen and charged heavy particles, $\nu_{{\mathrm {H}}}=\nu_{{\mathrm {H^+H}}}+\nu_{{\mathrm {He^+H}}}$. The values of temperature and number densities are taken from FAL93-3 model (Fontenla et al. \cite{Fontenla1993}). The lower panel shows the damping rates for the temperature of 8100 K with proton, neutral hydrogen and neutral helium number densities of $1.33 \times 10^{11}$ cm$^{-3}$,  $3.71 \times 10^{11}$ cm$^{-3}$ and $4.8 \times 10^{10}$ cm$^{-3}$, respectively. The middle panel shows the damping rates for the temperature of 10 000 K with proton, neutral hydrogen and neutral helium number densities of $1.07 \times 10^{11}$ cm$^{-3}$,  $8.47 \times 10^{10}$ cm$^{-3}$ and $1.13 \times 10^{10}$ cm$^{-3}$, respectively. The upper panel shows the damping rates for the temperature of 16 000 K with proton, neutral hydrogen and neutral helium number densities of $9.09 \times 10^{10}$ cm$^{-3}$,  $1.05 \times 10^{10}$ cm$^{-3}$ and $2.5 \times 10^{9}$ cm$^{-3}$, respectively. The three different temperatures correspond to 1790 km, 1995 km and 2015 km heights from the photosphere.

According to Fig. 2 the influence of neutral helium atoms on the damping of Alfv\'en waves significantly depends on the plasma temperature. The effect of neutral helium is less pronounced for the temperature of $\sim$ 8000 K (lower panel). But it gradually increases for higher temperatures (middle and upper panels). The dependence of the damping rate on frequency is similar to that for the neutral hydrogen: the damping rate increases at lower frequencies, reaches its maximum at particular frequency and then begins to decrease for higher values (Zaqarashvili et al. \cite{Zaqarashvili2011}). The damping rate of Alfv\'en waves has its maximum near ion-neutral collision frequency.

The results show that the neutral helium atoms significantly enhance the damping of Alfv\'en waves in the chromospheric plasma. For example, the upper panel of Fig. 2 corresponds to the parameters of spicules with the temperature of 16000 K (Beckers \cite{Beckers1968}). The presence of neutral helium enhances the damping rate almost three times for low-frequency waves and almost twice for high-frequency ones. Therefore, neutral helium atoms should be taken into account, alongside with neutral hydrogen, in the study of transverse waves in solar spicules (Zaqarashvili and Erd{\'e}lyi \cite{Zaqarashvili2009}).

\begin{figure}
\begin{center}
\includegraphics[width=8cm]{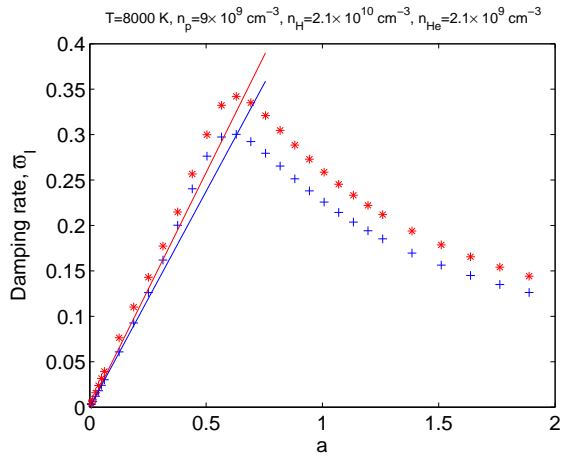}
\end{center}
\caption{Damping rate (imaginary part of frequency) of Alfv\'en waves, ${\varpi}={{\omega}/ {k_zv_A}}$, vs normalized Alfv\'en frequency, $a={{k_z
v_A}/ {\nu_{{\mathrm {H}}}}}$, for the prominence plasma. Red asterisks (blue crosses) correspond to the solutions with (without) neutral helium. Red (blue) solid line corresponds to the damping rate derived in the single-fluid approach (Eq. \ref{single}) with (without) neutral helium.}
\end{figure}

Recently, Soler et al. (\cite{Soler2010}) have studied the damping of MHD waves due to ion-neutral collision in the single-fluid description of prominence plasma including neutral helium atoms. They concluded that the neutral helium has no significant influence on the damping of MHD waves. We check the result in the three-fluid MHD description taking plasma conditions typical for prominence cores with 8000 K temperature. Proton and neutral hydrogen number densities are taken as $9 \times 10^{9}$ cm$^{-3}$ and  $2.1 \times 10^{10}$ cm$^{-3}$, respectively (Labrosse et al. \cite{Labrosse2010}). The number density of neutral helium is taken as 10 $\%$ of the neutral hydrogen i.e. $2.1 \times 10^{9}$ cm$^{-3}$ and the number density of singly ionized helium is taken as $2.1 \times 10^{8}$ cm$^{-3}$. Fig. 3 displays the damping rates of Alfv\'en waves in partially ionized isothermal plasma with (red asterisks) and without (blue crosses) neutral helium.  We see that the neutral helium has insignificant effect on the wave damping in considering parameters, especially for low-frequency waves, as it was shown by Soler et al. (\cite{Soler2010}). However, the neutral helium can be of importance in prominence-to-corona transition regions, where the plasma temperature is higher than in cores. Solid lines on this figure indicate the damping rates in the single-fluid approach. The damping rates derived in single- and multi-fluid approaches are similar for low-frequency waves, but have completely different behaviour when wave frequency approaches ion-neutral collision frequency.

\section{Discussion}

Solar chromosphere/photosphere and prominences contain significant amount of neutral atoms, which may lead to the damping of MHD waves. In some cases, the damping can be of importance and should be taken into consideration. The effect of neutral hydrogen on the damping of MHD waves is well studied in both, single fluid (Khodachenko el al. \cite{Khodachenko2004}, Forteza et al. \cite{Forteza2007}) and multi-fluid approximation (Zaqarashvili et al. \cite{Zaqarashvili2011}). However, the collision of ions with neutral helium atoms for some cases is important as shown here. Hydrogen atoms quickly become ionized with increasing temperature, while the number of neutral helium may be still significant. This is clearly seen in Fig. 4, which is plotted according to the atmospheric model FAL93-3. The ratio of neutral helium to neutral hydrogen number densities grows significantly in the temperature interval $10 000-40 000$ K.

We study the damping of Alfv\'en waves due to collision of ions with neutral hydrogen and neutral helium atoms. The three-fluid description of plasma is used, where one component is electron-proton-singly ionized helium and the other two components are neutral hydrogen and neutral helium gases. The dynamics of linear Alfv\'en waves in homogeneous isothermal plasma with constant unperturbed magnetic field is considered. The exact dispersion relation for the Alfv\'en waves is derived and solved for different plasma parameters. The damping rates due to ion collision with neutral hydrogen and neutral helium atoms are derived and compared with those obtained only for neutral hydrogen. The analysis shows that the collision between ions and neutral helium can be of importance at certain values of plasma temperature, when the hydrogen already begins ionization, but the neutral helium is still presented. This happens for $T\sim 10000-40000$ K, which corresponds to upper chromosphere, spicules and prominence-corona transition regions. The presence of neutral helium enhances the wave damping by few times compared to the damping only due to ion-neutral hydrogen collisions (Fig. 2, middle and upper panels).

\begin{figure}
\begin{center}
\includegraphics[width=8cm]{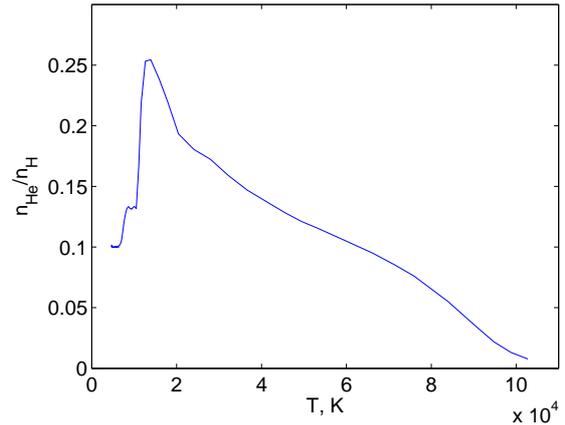}
\end{center}
\caption{The ratio of neutral helium and neutral hydrogen number densities, $n_{{\mathrm {He}}}/n_{\mathrm {H}}$ vs plasma temperature according to FAL93-3 model (Fontenla et al. \cite{Fontenla1993}).}
\end{figure}

The collisions between ions and neutral helium atoms can be of importance in solar spicules which correspond to the upper panel on Fig. 2. The plasma temperature, electron density and ionization degree are almost constant along spicules (especially in their upper part), therefore the isothermal plasma is a good approximation there  (Beckers \cite{Beckers1968}).
The neutral helium atoms may significantly enhance the damping of transverse waves, which are continuously observed in spicules (Kukhianidze et al. \cite{Kukhianidze2006}, De Pontieu et al. \cite{De Pontieu2007}, Zaqarashvili et al. \cite{Zaqarashvili2007}, see also recent review of Zaqarashvili and Erd{\'e}lyi \cite{Zaqarashvili2009}).

The presence of neutral helium has no significant influence on Alfv\'en wave damping in prominence cores (Fig. 3). This makes still valuable the results of Soler et al. (\cite{Soler2010}), who studied the damping of MHD waves in single-fluid approach. However, the neutral helium can be of importance in the prominence-corona transition region, where the plasma temperature is higher than in the prominence cores.

Damping rates derived in the three-fluid approach reach a peak near the ion-neutral collision frequency and then decrease for higher frequencies unlike to the single-fluid approach, where the damping linearly increases with increasing frequency (solid lines on Figs. 2-3, see also Zaqarashvili et al. \cite{Zaqarashvili2011}). This may require the reconsideration of high-frequency Alfv\'en wave damping in the chromosphere. More sophisticated simulations taking into account the gradients of temperature and ionization degrees with height in the different parts of solar atmosphere are needed to be performed.

In order to understand the wave damping due to ion-neutral collision, we consider the simplest case of partially ionized plasma consisting of electrons, protons and neutral hydrogen. Then, the well-known expression of damping rate in the single-fluid approach (Braginskii \cite{Braginskii1965}, Khodachenko el al. \cite{Khodachenko2004}, Forteza et al. \cite{Forteza2007}) can be rewritten as
\begin{equation}\label{single}
{\varpi_i}={{\omega_i}\over {k_zv_A}}={{k_zv_A \rho_{\mathrm {tot}}}\over {\alpha_{in}}}{{\xi^2_n}\over {2}}={{k_zv_A}\over {\nu_{{in}}}}{{\xi_n}\over {2\xi_i}},
\end{equation}
where $\rho_{\mathrm {tot}}$ is the total density (ion+neutral) and $\nu_{{in}}$ is the ion-neutral collision frequency defined by Eq. (\ref{nu_in}), ${\xi_n},{\xi_i}$ are normalised by $\rho_{\mathrm {tot}}$. This expression clearly indicates that the normalised damping rate depends on the ratio of Alfv\'en and ion-neutral collision frequencies and the ratio of neutral and ion fluid densities. Early statement that the damping rate depends on the magnetic field strength can be translated as follows: the increase of the magnetic field leads to the increase of the Alfv\'en speed, therefore the waves with the same wavenumber have higher frequency, which is closer to ion-neutral collision frequency, hence the normalised damping rate increases. In fact, it is the ratio of neutral and ion fluid densities, which determines the damping rate of particular wave harmonic. High-frequency waves are damped quickly. However, this statement is valid for low-frequency wave spectrum below ion-neutral collision frequency.

FAL93-3 model does not include the coronal values of plasma parameters and the ionization degree of helium in the corona is unknown.
However, it is clear that the coronal plasma contains only a small number of neutral helium, therefore the effects of neutral helium should be smaller in the corona. However, certain influence of neutral helium on wave damping in the corona is not ruled out, and further analysis is required to make a firm conclusion.

\section{Conclusions}

Existence of neutral helium atoms, alongside with neutral hydrogen, significantly enhances the damping of Alfv\'en waves at certain interval of plasma temperature due to ion-neutral collision as compared to the case of only neutral hydrogen. Neutral helium causes the increase of damping rates few times in spicules and prominence-corona transition region at the temperature $T \sim 10000-40000$ K.

Alfv\'en waves have maximal damping rates at some frequency interval peaking near the ion-neutral collision frequency. At the same time, the damping rate is reduced for higher frequencies.

The study of high-frequency Alfv\'en wave damping in the solar chromosphere with realistic height profile of ionization degree needs to be revised in the future including both, neutral hydrogen and neutral helium atoms.

\begin{acknowledgements}
The work was supported by the Austrian Fonds zur F\"orderung
der wissenschaftlichen Forschung (project P21197-N16). T.V.Z. also acknowledges
financial support from the Georgian National Science Foundation (under grant GNSF/ST09/4-310). The authors thank to the referee
for constructive comments.
\end{acknowledgements}

\appendix

\end{document}